\documentclass[aps,pra,showpacs,twocolumn,nofootinbib]{revtex4-2}
\usepackage[colorlinks=true, allcolors=blue]{hyperref}
\usepackage{bm}
\usepackage{amsmath}\allowdisplaybreaks
\usepackage{amssymb}
\usepackage{slashed}
\usepackage{float}
\usepackage{relsize}

\usepackage{subcaption}
\usepackage{dcolumn}
\usepackage{graphicx}
\usepackage{xcolor}
\usepackage{nicefrac}
\usepackage{physics}
\usepackage{multirow}
\usepackage{booktabs}

\newcolumntype{.}{D{x}{}{-1}}

\newcolumntype{w}[1]{D{.}{.}{#1}}
\newcolumntype{L}{>{$}l<{$}}

\newcommand{\Za}{{Z\alpha}}

\newcommand{\lbr}{\langle}
\newcommand{\rbr}{\rangle}

\begin{document}

\title{Model-independent determination of nuclear charge radii from Li-like ions}

\author{V.~A. Yerokhin}
\affiliation{Max~Planck~Institute for Nuclear Physics, Saupfercheckweg~1, D~69117 Heidelberg, Germany}

\author{Ben Ohayon}
\affiliation{Physics Department, Technion—Israel Institute of Technology, Haifa 3200003, Israel}

\begin{abstract}

We demonstrate that recent advances in QED theory of Li-like ions [V.~A.~Yerokhin {\em et al.}, Phys. Rev. A {\bf 112}, 042801 (2025)]  enable determinations of absolute nuclear charge radii for heavy elements.
By incorporating constraints derived from electron-scattering data, we obtain radii that are independent of the assumed model of the nuclear charge distribution.
Our approach is validated for $^{208}$Pb, a well-studied spherical nucleus, and is then applied to $^{209}$Bi, where low-lying nuclear excitations complicate the interpretation of muonic–atom data.

\end{abstract}

\maketitle

\section{Introduction}

The nuclear root-mean-square (rms) charge radius is a fundamental parameter that determines the deviation of the electrostatic interaction between the nucleus and atomic electrons from its pure Coulomb form.
Its accurate knowledge is important for tests of bound-state QED~\cite{indelicato:19}, understanding of nuclear beta-decay~\cite{2024-SG,Gorchtein:25},
determining the nuclear neutron-distribution radii and properties of neutron-star crusts~\cite{horowitz:01,abrahamyan:12},
as well as for studies of parity violation in atomic systems~\cite{Wansbeek:08}.
Since accurate {\em ab initio} calculations of nuclear radii are currently not feasible, the values available in the literature are determined from experiment, primarily from electron–nucleus scattering~\cite{vries:87} and spectroscopy of muonic atoms~\cite{fricke:04}.

Extracting nuclear radii from electron scattering data is complicated by the need to extrapolate to zero momentum transfer and requires a thorough understanding of numerous relativistic and radiative corrections~\cite{sick:18}.
For nuclei with nonzero spin, it is also necessary to separate the electric and magnetic form factors.
Due to these difficulties, the typical deviations between nuclear radii derived from the electron–scattering and muonic–atom measurements are on the level of $1\%$ or worse~\cite{Fricke:92}, which is generally attributed to systematic uncertainties in the electron–scattering analysis.

Nuclear radii derived from muonic atoms are typically regarded as more reliable; however, their accuracy is limited by incomplete knowledge of the nuclear structure, specifically the nuclear polarization and the shape of the nuclear charge distribution.
A more general analysis method~\cite{fricke:95} combines electron–scattering data with muonic–atom energies.
In this approach, the measured energies are first converted into the so-called Barrett radii~\cite{barrett:70} and then corrected for the nuclear shape determined from electron–scattering data.
This procedure delivers nuclear radii accurate to about 0.1\% in the medium-$Z$ range~\cite{ohayon:25:radii}.
A comparable level of precision has recently been achieved in a new generation of purely muonic–atom
determinations~\cite{sun:25}.

Another promising way to access the nuclear radii is through the spectroscopy of {\em electronic} atoms.
Accurate results have been reported for the {\em differences} of nuclear radii between isotopes of the same element~\cite{nortenhauser:11,nortenhauser:15,werf:25,katyal:25}.
However, determinations of the {\em absolute} values of nuclear radii from electronic atoms have so far been accomplished only for hydrogen and deuterium~(see \cite{mohr:25:codata} and references therein). The main limitation has been
the difficulty of accurately describing systems with more than one electron in theoretical calculations.
However, recent advances in {\em ab initio} QED theory of few-electron ions~\cite{malyshev:23,yerokhin:25:sese,malyshev:25:belike:tobe}
have now made such determinations feasible. Nuclear radii can be extracted not only from the transition energies but also from the bound-electron $g$ factors.
In particular, the recent $g$-factor measurement of H-like tin~\cite{morgner:23}, combined with advanced QED calculations~\cite{sikora:25}, allows determining the $^{118}$Sn charge radius with an accuracy of about 0.2\%.

The $2p_j\,$--$\,2s$ transitions in Li-like ions are excellent candidates for extracting the charge radii of high-$Z$ nuclei, as they offer an attractive compromise between the feasibility of high-precision QED calculations and the  practicality of accurate experimental measurements. The latest QED calculations of Li-like ions \cite{yerokhin:25:lilike} have opened the way to determining the nuclear radii of high-$Z$ elements with an accuracy at the  0.2\% level, with further improvements  expected in the near future.

To extract absolute nuclear radii from the spectra of electronic atoms, it is essential to quantify the uncertainty associated with the nuclear charge-distribution model employed in theoretical calculations.
In most previous studies, this uncertainty was relatively small compared to other theoretical and experimental errors, and it was typically estimated by comparing results obtained with two models: the Fermi distribution and the homogeneously charged sphere model~\cite{johnson:85,franosch:91,yerokhin:15:Hlike}.
Naturally, such a simple approach is not sufficient for determining the nuclear radius.

The goal of the present work is to formulate an approach for a model-independent determination of nuclear charge radii from the measured spectra of electronic ions. To this end, we study the dependence of the finite nuclear size correction on the nuclear model and show that this dependence can be eliminated by constraining the models with the available electron-scattering data.
We will demonstrate the procedure by determining the nuclear charge radii of two high-$Z$ elements, $^{208}$Pb and $^{209}$Bi, using the existing spectroscopic data for the $2p_j\,$--$\,2s$ transition energies of the corresponding Li-like ions.

\section{Finite nuclear size correction}

The spatial charge distribution of the nucleus is described by the electric form factor $\rho(q)$, where $q$ is the momentum transfer in a scattering process.
The low momentum expansion of $\rho(q)$ takes the form
$$
\rho(q)=1-\frac{q^2}{6}\langle r^2\rangle+\frac{q^4}{120}\langle r^4\rangle-\frac{q^6}{5040}\langle r^6\rangle+\cdots,
$$
where $q^n$ are accompanied by the corresponding $n$th moment of the radial charge distribution. The model-independent moments $\lbr r^{2k}\rbr$ thus describe the large-$r$ shape of the nuclear charge distribution but do not provide much information about the small-$r$ (large-$q$) region.

The finite nuclear size (fns) correction to energy levels has been extensively studied within the $\Za$ expansion. In relativistic units, it is given by
\cite{pachucki:18,pachucki:25:recfns}
\begin{align} \label{eq:fns}
E_{\rm fns} = &\ \frac{2\pi\Za }{3}\, \phi^2(0)\,r_C^2
 - \frac{\pi(\Za)^2}{3}\, \phi^2(0)\,r_F^3
 \nonumber \\ &
 - \frac{2\pi(\Za)^3 }{3}\, \phi^2(0)\,r_C^2\,\ln \left( \Za\, r_C\right)
 + \ldots \,,
\end{align}
where $\phi^2(0)$
is the electron charge density at $r = 0$, $r_C \equiv \langle r^2\rangle^{1/2}$ is the rms nuclear charge radius, $r_F$ is the Friar radius~\cite{friar:79:ap} defined by
$$
r_F^3 = \int d^3r_1d^3r_2\,\rho(r_1)\,\rho(r_2)\,|\vec{r}_1-\vec{r}_2|^3\,,
$$
and $\ldots$  denote higher-order contributions.
We note that the fns correction $E_{\rm fns}$ depends on the shape of the nuclear charge only through some integral quantities, first of all, $r_C^2$.
Higher-order terms in Eq.~(\ref{eq:fns}) contain further momenta of nuclear charge distribution: $r_F$, $r_C^4$, $\langle r^4\rangle$, $\kappa_1$, $\kappa_2$,
see Refs.~\cite{pachucki:18,pachucki:25:recfns} for details.
It should be emphasized that, although the nuclear charge form factor can be expressed as an expansion in $\langle r^{2k}\rangle$, this does not imply that the fns correction $E_{\rm fns}$ can be expanded in the same manner.

For high-$Z$ ions, the $\Za$ expansion formulas are no longer applicable, and the dependence of the fns correction on nuclear parameters is less transparent.
It is known~\cite{shabaev:93:fns} that the leading dependence of the fns correction on $r_C$, which is $\propto r_C^2$ in the nonrelativistic limit, changes to $\propto r_C^{2\gamma}$ in the relativistic case, with $\gamma = \sqrt{1-(\Za)^2}$. Consequently, the power of $r_C$ deviates significantly from 2 in the high-$Z$ region.

\section{Model dependence of nuclear-size correction}
\label{sec:model}
\begin{table*}
\caption{
Finite nuclear size correction to the energy of the $1s$ state of H-like lead ion ($Z = 82$)
calculated with different models of the nuclear charge distribution, with $r_C = 5.5030~\mbox{\rm fm}$.
$n$ denotes the number of constrained nuclear-charge moments,
``Parameters'' lists the parameters of the model (in fermi),
$E_{\rm fns}$ denotes the fns energy corrections (in eV),
``Diff.'' denotes the deviation from the Fermi model with the same $n$.
}
\label{tab:FNS}
\begin{ruledtabular}
\begin{tabular}{ccccccccccr}
\multicolumn{1}{c}{$n$} &
        \multicolumn{1}{c}{Model} &
          \multicolumn{1}{c}{Eq.} &
            \multicolumn{1}{c}{Parameters} &
                \multicolumn{1}{c}{$r_C$} &
                    \multicolumn{1}{c}{$V_{24}$} &
                        \multicolumn{1}{c}{$V_{26}$} &
                            \multicolumn{1}{c}{$V_{28}$} &
                            \multicolumn{1}{c}{$r_F/r_C$} &
                                \multicolumn{1}{c}{$E_{\rm fns}$} &
                                    \multicolumn{1}{r}{Diff.}
                                       \\
\hline\\[-5pt]
1 & 1pF & (\ref{eq:1pF}) & $c = 6.64554$      & 5.5030 & 0.9399 & 0.8959 & 0.8600 & 1.4912 & 67.21373 \\
  & 1pH & (\ref{eq:1pH}) & $ R = \sqrt{5/3}\,r_C$     & 5.5030 & 0.9573 & 0.9302 & 0.9112 & 1.4856 & 67.34240 &  0.19\% \\
  & 1pG & (\ref{eq:1pG}) & $ R = r_C$ & 5.5030 & 0.8801 & 0.7974 & 0.7356 & 1.5146 & 66.66235 & $-$0.82\% \\
  & 1pE & (\ref{eq:1pE}) & $ R = r_C$ & 5.5030 & 0.7953 & 0.6640 & 0.5718 & 1.5590 & 65.63333 & $-$2.35\% \\ \hline\\[-7pt]
2 & 2pF & (\ref{eq:2pF}) & $\mathsmaller{c = 6.6652,\ a = 0.51242}$  & 5.5030 & 0.9406 & 0.8974 & 0.8621 &  1.4910 & 67.21939 \\
  & 2pH & (\ref{eq:npH}) & $\mathsmaller{C_1 = 1/4,\ R_1 = 5.17961,\ R_2 = 7.6389}$ & 5.5030 & 0.9406 & 0.9028 & 0.8768 &  & 67.21107 & $-$0.012\% \\
  & 2pG & (\ref{eq:npG}) & $\mathsmaller{C_1 = 1/4,\ R_1 = 3.15548,\ R_2 = 5.68309}$& 5.5030 & 0.9406 & 0.9041 & 0.8772 & 1.4912 &  67.20764 & $-$0.018\% \\
  & 2pE & (\ref{eq:npE}) & $\mathsmaller{C_1 = 1/5,\ R_1 = 4.68623,\ R_2 = 5.2547}$ & 5.5030 & 0.9406 & 0.8763 & 0.7884 & 1.4905 & 67.24369 &  0.036\% \\ \hline\\[-7pt]
3 & 3pF & (\ref{eq:3pF}) & $\mathsmaller{c = 6.81571,\ a = 0.4969,\ w=-0.157696}$ & 5.5030 & 0.9406 & 0.8983 & 0.8645 & 1.4910 & 67.21806 & \\[5pt]
  & 3pH & (\ref{eq:npH}) & ${C_1 = 1/4,\ C_2 = 1/2,}\atop{ R_1 = 5.9286,\ R_2 = 6.86233,\ R_3 = 8.51794}$        & 5.5030 & 0.9406 & 0.8983 & 0.8658 & & 67.21832 & 0.0004\% \\[5pt]
  & 3pG & (\ref{eq:npG}) & ${C_1 = 1/3,\ C_2 = 1/3,}\atop{ R_1 = 4.49451,\ R_2 = 4.90557,\ R_3 = 6.0888}$        & 5.5030 & 0.9406 & 0.8983 & 0.8640 & 1.4910 & 67.21788 & $-$0.0003\% \\[5pt]
  & 3pE & (\ref{eq:npE}) & ${C_1 = 1/6,\ C_2 = 1/2,}\atop{ R_1 = 3.53028,\ R_2 = 4.76646,\ R_3 = 6.21784}$       & 5.5030 & 0.9406 & 0.8983 & 0.8618 & 1.4910 &  67.21716 & $-$0.0013\% \\[5pt]
\end{tabular}
\end{ruledtabular}
\end{table*}
In this section, we study the model dependence of the fns correction to energy levels.
To this end, we perform a series of calculations with different models of the nuclear
charge distribution,  all constrained by three model-independent parameters: $r_C$, $V_{24}$, and $V_{26}$, where
\begin{align}
V_{24} = \frac{r_C}{\langle r^4\rangle^{1/4}}\,,
\ \ V_{26} = \frac{r_C}{\langle r^6\rangle^{1/6}}\,.
\end{align}
We prefer to use the ratios $V_{24}$ and $V_{26}$ rather than the absolute moments $\langle r^4\rangle^{1/4}$ and $\langle r^6\rangle^{1/6}$, because the ratios can be more accurately extracted from the electron scattering data (see Ref.~\cite{fricke:95} and the the discussion in the next section).
The spread of energy shifts obtained with different models will serve as an estimate of model dependence.
Specifically, we will adopt the parameter set
\begin{align} \label{eq:set}
r_C = 5.5030~\mbox{\rm fm}, \quad V_{24} = 0.9406, \quad V_{26} = 0.8983,
\end{align}
which corresponds to the $^{208}$Pb isotope with $Z = 82$.
Note that the actual numerical values of the parameters are not essential for us at this stage; our objective is to assess the model dependence of our calculations while
keeping the nuclear parameters fixed.

We consider three classes of models for the nuclear charge distribution.
The first class consists of one-parameter models:
Fermi (1pF), homogeneous (1pH), Gaussian (1pG), and exponential (1pE),
see Appendix~\ref{sec:models} for details.
With one-parameter models, only a single nuclear parameter can be reproduced
(naturally, it is chosen to be $r_C$).
The second class consists of two-parameter models:
Fermi (2pF), homogeneous (2pH), Gaussian (2pG), and exponential (2pE),
defined in Appendix~\ref{sec:models}.
These models allow us to fit two nuclear parameters, $r_C$ and $V_{24}$.
The third class contains models with three free parameters:
Fermi (3pF), homogeneous (3pH), Gaussian (3pG), and exponential (3pE), see Appendix~\ref{sec:models}.
In this case, three nuclear parameters, $r_C$, $V_{24}$, and $V_{26}$, can be fitted.

The results of our numerical calculations are summarized in Table~\ref{tab:FNS}.
The first column displays the number of fitted nuclear moments, $n$.
The fourth column lists the numerical values of the model parameters used in our calculations. These parameters were
obtained by fitting the nuclear moments from Eq.~(\ref{eq:set}).
The next five columns present  values of the charge distribution moments computed for each model. The rms radius is, of course, identical across all models by construction.
The $V_{24}$ parameter is the same for models with $n\ge2$, but  different in the $n=1$ case. $V_{26}$ is identical for models with $n=3$. The last two columns
present the numerical values of the fns correction for the $1s$-state energy and the deviations from the Fermi-model value with the same $n$.

We find that when only the rms radius is fixed, various models yield fns corrections that differ already at the $1\%$ level. By additionally constraining the next-order moment, $\langle r^4 \rangle$, the spread of the fns corrections is reduced by nearly two orders of magnitude to approximately $0.02\%$.
Finally, when all three lowest moments of the nuclear charge distribution, $\langle r^2 \rangle$, $\langle r^4 \rangle$, and $\langle r^6 \rangle$, are fixed, the spread is further suppressed to a practically negligible level of $0.001\%$.

Table~\ref{tab:FNS} also presents the numerical values of the ratio $r_F/r_C$ obtained with different models. The Friar moment $r_F$ originates from a high-momentum exchange region, and so far, it has been considered an independent nuclear parameter.
It came as a surprise to us that the Friar moment also becomes practically model-independent when the first three moments of the charge distribution are constrained.

Summarizing our findings presented in Table~\ref{tab:FNS}, we conclude that once the first three moments of the nuclear charge distribution are known, numerical calculations of the fns correction can be performed in an almost model-independent way, with the residual model dependence reduced to the $10^{-5}$ level, which is completely negligible at the present level of interest.

\begin{figure*}
\centerline{
\resizebox{\textwidth}{!}{%
  \includegraphics{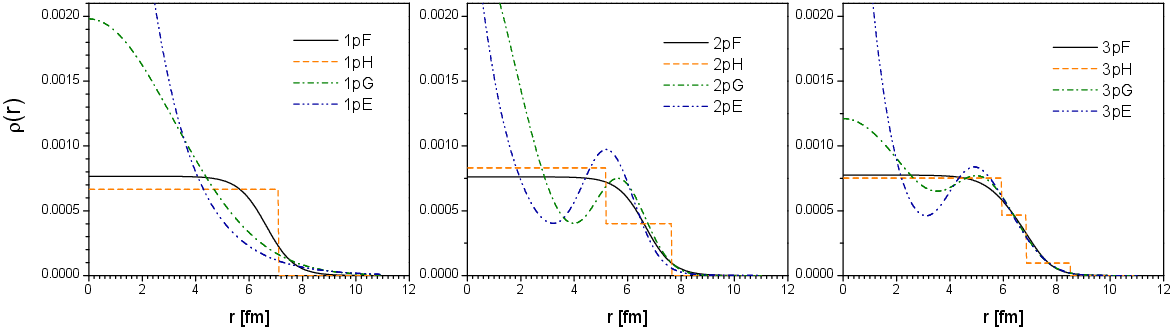}
}}
 \caption{
 Nuclear charge distributions for $^{208}$Pb with $r_C = 5.5030$~fm
 in various models: 1-parameter models (left graph),
 2-parameter models (middle graph), 3-parameter models (right graph).
 Model parameters are specified in Table~\ref{tab:FNS}.
\label{fig:distr}}
\end{figure*}

In Figure~\ref{fig:distr}, we plot the radial dependence of the nuclear charge distributions summarized in Table~\ref{tab:FNS}. We observe that as the number of constrained moments increases, the distributions tend to converge in the outer nuclear region  ($r \gtrsim r_C$). In contrast, no such convergence is observed in the inner region ($r \ll r_C$). This suggests that the form of the charge distribution at small radial distances has very little impact on the fns correction. It is also instructive to see that the visually very different distributions in the right panel of the figure yield essentially the same fns corrections.

We conclude that a ``realistic" shape of the nuclear charge distribution is not required for an accurate determination of the fns correction to energy.
In particular, the $n$pH distribution defined in Eq.~(\ref{eq:npH}), while clearly not very realistic, yields the fns correction that is essentially as accurate as that obtained from a more physical Fermi distribution, provided that the three leading moments of the charge distribution are constrained.

It should be mentioned that although our analysis in Table~\ref{tab:FNS} is carried out only for the $1s$ hydrogenic state, the conclusions should be applicable also
for transitions in many-electron atoms.
The reason is that the dominant fns effect in atomic transition energies comes from the
mixing of the valence electrons with core $s$-shell electrons by the electron-electron
interaction.

\section{Determination of \texorpdfstring{$V_{24}$}{V24} and \texorpdfstring{$V_{26}$}{V26} from electron-scattering data}

In the previous section, we established that a detailed description of the nuclear shape is not required for accurate radius extractions; only the ratios $V_{24}$ and $V_{26}$ are needed.
These ratios may be estimated using nuclear theory~\cite{Beyer:25, miyagi2025ab} or obtained through elastic electron scattering~\cite{vries:87}. Here, we explore the latter approach.

Many high quality scattering experiments have been performed on $^{208}$Pb~\cite{1969-Heisenberg, 1969-Niftrik, 1971-Nagao, 1972-Friedrich, 1976-Euteneuer, 1977-Frois, 1978-Euteneuer}, a doubly-magic, spherical heavy nucleus.
The data has been reanalyzed numerous times, also using so-called ``model-independent" methods, i.e., those that do not assume a particular charge distribution.
Three such analyses are based on the Fourier-Bessel method~\cite{1978-Euteneuer, 1977-Pb, fricke:95}, and one employs the sum-of-Gaussians method~\cite{1977-Frois}.
In three of these studies, the low momentum transfer limit has been constrained by data  from muonic-atom spectroscopy, as detailed in Ref.~\cite{vries:87}.

Table~\ref{tab:V} presents  the moments and their ratios extracted from various $^{208}$Pb charge distributions reported in the literature. It is remarkable that the individual moments are found to differ by up to $0.5\%$, depending on the chosen data set, analysis method, and the inclusion of muonic-atom results. However, their ratios are an order of magnitude more stable. Moreover, these ratios are expected to be less susceptible to systematic uncertainties, such as those stemming from normalization~\cite{fricke:95}.

For the purpose of extracting charge radii using $V_{24}$ and $V_{26}$, we adopt the average of the individual results as the recommended values, with the standard deviation taken as an estimate of the uncertainty.
For $^{209}$Bi, the available data are insufficient to estimate the uncertainty directly; therefore, we assign  the same error as for $^{208}$Pb due to the similarity of the experimental results.

\begin{table*}[htbp]
\centering
\begin{ruledtabular}
\caption{
Moments and ratios $V_{24}$ and $V_{26}$ of charge distribution in $^{208}$Pb and $^{209}$Bi.
The five first columns contain, from left to right: the isotope, the main reference(s), the method of analysis (FB-Fourier-Bessel, SOG-Sum of Gaussians), whether muonic atoms have been used as a constraint (V-yes, X-no),
the range of transferred momenta $q$ included in the analysis.
Due to the similarity in the data and analysis, the same uncertainty is assumed for $V_{24}$ and $V_{26}$
in $^{209}$Bi
as for $^{208}$Pb.\label{tab:V}
}
\label{table:nparam}
\begin{tabular}{llll lll lll}
 Nucleus & Ref. & Method & $\mu$ & q-range & $\langle r^2\rangle^{1/2}$ & $\langle r^4\rangle^{1/4}$ & $\langle r^6\rangle^{1/6}$ & $V_{24}$ & $V_{26}$\\
\midrule
$^{208}$Pb & \cite{1978-Euteneuer, vries:87} & FB & V & 0.5$-$2.2& 5.5032  & 5.8505 & 6.1279 & 0.9406 & 0.8981 \\
$^{208}$Pb & \cite{fricke:95} & FB & X & 0.5$-$2.2& 5.4785  & 5.8224 & 6.0909 & 0.9409 & 0.8995 \\
$^{208}$Pb & \cite{1977-Pb, vries:87} & FB & V & 0.4$-$3.7& 5.4993  & 5.8427 & 6.1123 & 0.9412 & 0.8997 \\
$^{208}$Pb & \cite{1977-Frois, vries:87} & SOG& V & 1.7$-$3.7& 5.5003  & 5.8464 & 6.1208 & 0.9408 & 0.8986 \\
$^{208}$Pb &\multicolumn{3}{l}{Average(Std.)}&&
5.495(11) & 5.841(12) & 6.113(16) & 0.9409(2) & 0.8990(8) \\
$^{208}$Pb &\cite{miyagi2025ab} & Theory &&&&&& 0.9404 &  \\

\\
$^{209}$Bi & \cite{1978-Euteneuer, vries:87} & FB & V & 0.5$-$2.2 & 5.5184 & 5.8606 & 6.1284 & 0.9416(2) & 0.9004(8) \\
\end{tabular}
\end{ruledtabular}
\end{table*}
%
\section{Nuclear radii determination from Li-like transition energies
}

We now turn to the determination of nuclear charge radii from the $2p_j\,$–$\,2s$ transition energies in Li-like ions.
High-precision theoretical predictions for these transitions were recently reported in
Ref.~\cite{yerokhin:25:lilike} for the entire isoelectronic sequence with $Z \ge 10$.
Theoretical energies become increasingly sensitive to the
nuclear radii as the nuclear charge $Z$ increases,
thus making high-$Z$ ions promising candidates for a competitive determination of nuclear radii.
Among the heavy elements, $^{208}$Pb and $^{209}$Bi are currently the only cases for which
both electron-scattering data and precise spectroscopic measurements of the relevant transition
energies are available.
In the following, we extract the nuclear charge radii for these two nuclei.

\subsection{Charge radius of \texorpdfstring{$^{208}$Pb}{}}

There are two experimental results
for $^{208}$Pb, one for the $2p_{1/2}$-$2s$ ($D_1$)
transition~\cite{brandau:03}
\begin{align}\label{eq:Pb:exp:D1}
E_{\rm exp}(D_1) = 230.65\,(8)~\mbox{\rm eV}\,,
\end{align}
and another for the $2p_{3/2}$-$2s$ ($D_2$)
transition~\cite{zhang:08}
\begin{align}\label{eq:Pb:exp:D2}
E_{\rm exp}(D_2) = 2642.26\,(10)~\mbox{\rm eV}\,.
\end{align}
The theoretical transition energies are taken from Ref.~\cite{yerokhin:25:lilike},
\begin{align}\label{eq:Pb:the}
E_{\rm the}(D_1;r_{C,0},V_{24,0},V_{26,0}) = 230.817\,(31)~\mbox{\rm eV}\,,
 \nonumber \\
E_{\rm the}(D_2;r_{C,0},V_{24,0},V_{26,0}) = 2642.210\,(30)~\mbox{\rm eV}\,,
\end{align}
where we kept only the purely theoretical uncertainty and $r_{C,0}$, $V_{24,0}$, and $V_{26,0}$ are the nuclear parameters adopted in Ref.~\cite{yerokhin:25:lilike}. Specifically,
\begin{align}\label{eq:zero}
r_{C,0} = 5.5012~\mbox{\rm fm},\ V_{24,0} = 0.9398,\  V_{26,0} = 0.8959\,,
\end{align}
where $V$'s correspond to the standard one-parameter Fermi distribution model used in Ref.~\cite{yerokhin:25:lilike}.

We now form the field-shift difference
\begin{align}
\delta_{\rm fs}E  \equiv E_{\rm the}(r_C,V_{24},V_{26}) -
E_{\rm the}(r_{C,0},V_{24,0},V_{26,0})\,,
\end{align}
where $V_{24}$ and $V_{26}$ are the actual values of these parameters taken from Table~\ref{table:nparam}, and $r_C$ is a free parameter that needs to
be adjusted so that the equation
\begin{align} \label{eq:cond}
\delta_{\rm fs}E
 = E_{\rm exp} - E_{\rm the}(r_{C,0},V_{24,0},V_{26,0})
\end{align}
is fulfilled.

In order to determine $r_C$, we calculate the field-shift differences $\delta_{\rm fs}E$ numerically by slightly varying the parameters of our nuclear model around the initial point (\ref{eq:zero}) and recording the values of $\delta_{\rm fs}E$, $r_{C}$, $V_{24}$, and $V_{26}$ in each case.
We then fit the numerical results for $\delta_{\rm fs}E$ to the form
\begin{align} \label{eq:newseltz}
\delta_{\rm fs}E
 = C_2\,\delta \lbr r^2\rbr^{\gamma} + C_4\, \delta \lbr r^4\rbr + D_4\, \delta \lbr r^2\rbr^2 + C_6\, \delta \lbr r^6\rbr\,,
\end{align}
where $\delta \lbr r^{2n}\rbr = \lbr r^{2n}\rbr - \lbr r^{2n}\rbr_0$, $\gamma = \sqrt{1-(\Za)^2}$,
and the coefficients $C_n$ and $D_4$ are free parameters determined by fitting.
We note that the ansatz (\ref{eq:newseltz}) differs from
the one used by Seltzer \cite{seltzer:69}
in that we included $\gamma$ in the $C_2$ part and added the $D_4$ contribution.
This ansatz complies better with the analytical formula (\ref{eq:fns}) and
the high-$Z$ $r_C^{2\gamma}$ scaling. As a consequence, it yields a better
fit for our numerical data.
Nevertheless, we emphasize that Eq.~(\ref{eq:newseltz})
is not a rigorous analytical expansion but merely
an approximate representation.
As a consequence, the coefficients in Eq.~(\ref{eq:newseltz}) slightly depend on the reference point
(\ref{eq:zero}) and the chosen fitting range. We employ Eq.~(\ref{eq:newseltz}) only for interpolation of our numerical data
and have verified that the interpolation errors are fully negligible for our purposes.

In practical calculations, we computed a set of $\delta_\mathrm{fs}E$ values with the 3pF nuclear model,
and then fitted our results to the form (\ref{eq:newseltz}).
The analysis performed in Sec.~\ref{sec:model} shows that the actual choice of the nuclear model is irrelevant in this context.
In our calculations of $\delta_{\rm fs}E$, we included the Dirac energies, the one-photon exchange correction,
as well as the dominant parts of the two-photon exchange and one-loop QED effects; see Ref.~\cite{yerokhin:25:lilike} for details.

Specifically, our fit yields the coefficients
$C_2 = -0.662$~eV/fm$^2$, $C_4 = 3.22\times 10^{-4}$~eV/fm$^4$, $D_4 = -3.1\times 10^{-4}$~eV/fm$^4$, and $C_6 = -6.4 \times 10^{-7}$~eV/fm$^6$ for the $D_1$ transition,
as well as
$C_2 = -0.715$~eV/fm$^2$, $C_4 = 3.47\times 10^{-4}$~eV/fm$^4$, $D_4 = -3.4\times 10^{-4}$~eV/fm$^4$, and $C_6 = -6.9 \times 10^{-7}$~eV/fm$^6$
for the $D_2$ transition.

With the help of these coefficients, we obtain the following values for the
nuclear charge radius,
\begin{subequations} \label{eq:Pb:rms}
\begin{align}
r_C(^{208}\mathrm{Pb};D_1) = 5.558\,(27)_{\rm exp}(10)_{\rm the}\,\mathrm{fm}\,,
\label{eq:Pb:rms:a}
  \\
r_C(^{208}\mathrm{Pb};D_2) = 5.517\,(31)_{\rm exp}(9)_{\rm the}\,\mathrm{fm}\,,
\end{align}
\end{subequations}
from the $D_1$ and $D_2$ transitions, respectively.
Here, the first uncertainty arises from the experimental transition energy, while the second uncertainty originates from theory. The uncertainty due to errors in $V_{24}$ and $V_{26}$ is about $0.0001$~fm and thus completely negligible.

Since the two results in Eq.~(\ref{eq:Pb:rms}) agree within their combined statistical error, we may compute their weighted average, obtaining $5.540\,(20)_{\rm exp}(10)_{\rm the}\,\mathrm{fm}.$
It can be compared with the more accurate radius recently derived purely from muonic-atom spectroscopy, $5.5062\,(17)$~fm~\cite{sun:25}. The results are consistent with each other within $1.5$ times their combined standard uncertainty.
The small deviation from the more precise muonic-atom value could be due to the $D_1$ experimental energy.

\subsection{Charge radius of \texorpdfstring{$^{209}$Bi}{}}
For the determination of the nuclear charge radius of $^{209}$Bi, we employ the experimental result of Ref.~\cite{beiersdorfer:98} for the $2p_{3/2}-2s$ transition energy,
\begin{align}\label{eq:Bi:exp}
E_{\rm exp} = 2788.139\,(39)~\mbox{\rm eV}\,.
\end{align}
The theoretical transition energy is taken from Ref.~\cite{yerokhin:25:lilike},
\begin{align}\label{eq:Bi:the}
E_{\rm the}(r_{C,0},V_{24,0},V_{26,0}) = 2788.116\,(35)~\mbox{\rm eV}\,,
\end{align}
where only the purely theoretical uncertainty is retained, and the parameters $r_{C,0}$, $V_{24,0}$, and $V_{26,0}$ are the nuclear parameters adopted in Ref.~\cite{yerokhin:25:lilike}. Specifically,
\begin{align}
r_{C,0} = 5.5211~\mbox{\rm fm},\ V_{24,0} = 0.9400,\  V_{26,0} = 0.8962.
\end{align}

By calculating the field-shift differences $\delta_\mathrm{fs}E$ and fitting the numerical results to the form~(\ref{eq:newseltz}), we obtain
$C_2 = -0.810$~eV/fm$^2$, $C_4 = 3.90\times 10^{-4}$~eV/fm$^4$, $D_4 = -3.9\times 10^{-4}$~eV/fm$^4$, and $C_6 = -7.6 \times 10^{-7}$~eV/fm$^6$.
With the help of these coefficients, we obtain our result for the rms radius of $^{209}$Bi,
\begin{align} \label{eq:Bi:rms}
r_C(^{209}\mathrm{Bi}) = 5.514\,(11)_{\rm exp}(10)_{\rm the}\,\, \mathrm{fm}\,,
\end{align}
where the first uncertainty comes from the experimental transition energy (\ref{eq:Bi:exp}), and the second arises from the theoretical prediction (\ref{eq:Bi:the}).
Just as in the case of lead, the uncertainty from the errors in the $V$ factors is about $0.0001$~fm and thus negligible.

Comparing the rms radius of Eq.\,(\ref{eq:Bi:rms}) with literature values is not as straightforward as it is for $^{208}$Pb due to the hyperfine structure~\cite{Powers:68} and low-lying nuclear excitations~\cite{HUFNER:67, 1971-Rinker}.
Engfer \textit{et al.} quoted $r_C(^{209}\mathrm{Bi})=5.517(7)\,$fm~\cite{1974-Engfer} based on energies measured in Ref.~\cite{Powers:68}, in agreement with our result.
A more recent model-dependent analysis reported a different value, $r_C(^{209}\mathrm{Bi})=5.533(3)\,$fm~\cite{fricke:04}, where we added an error based solely on the experimentally measured energy. Our result of Eq.~(\ref{eq:Bi:rms}) indicates a slight preference for the radius quoted by Engfer.
A modern analysis of muonic $^{209}$Bi energy levels, similar to the one performed for $^{208}$Pb~\cite{sun:25}, could help clarify these deviations.
It is worth mentioning that extensive recent measurements are available~\cite{2024-Bi}.

\section{Conclusion and outlook}

In this work, we have shown that detailed knowledge of the nuclear charge distribution is not required for an accurate description of atomic energies, and that the first three moments of the nuclear charge distribution  -- $\lbr r^2\rbr^{1/2}$, $\lbr r^4\rbr^{1/4}$, and $\lbr r^6\rbr^{1/6}$ -- are sufficient for all practical purposes.
Using constraints on the ratios of these moments derived from electron-scattering experiments, we eliminated the dependence of theoretical energies on the shape of the nuclear charge distribution, thereby enabling model-independent determinations of absolute nuclear radii from atomic spectra.

The proposed approach is applied for the determination of nuclear charge radii of $^{208}$Pb and $^{209}$Bi from the $2p$-$2s$ transition energies in the corresponding Li-like ions. Our results are consistent with the radii derived from muonic-atom spectroscopy, although less accurate.
The main sources of our uncertainties are the experimental transition energies and the yet uncalculated two-loop QED effects, both of which can be significantly improved in the future.

It is important that the uncertainties due to the shape of the nuclear charge distribution (including those due to errors in the moments ratios $V_{24}$ and $V_{26}$)
are completely negligible in our approach. This is an advantage over the radii derived from muonic atoms, where the uncertainties due to the nuclear shape are significant and difficult to address reliably~\cite{ohayon:25:radii}.

Let us now discuss the improvements in the determination of nuclear radii that can be achieved in the foreseeable future.
For lead, the dominant source of uncertainty in the extracted radius originates from the experimental uncertainty of the $2p$–$2s$ transition energies, which is currently about $0.10$ eV. An experimental precision of $0.015$ eV has already been demonstrated for Li-like uranium~\cite{beiersdorfer:05}, indicating that a similar level of accuracy for lead and bismuth is within reach. Such precision in the transition energy would correspond to an uncertainty of approximately $0.005$ fm in the nuclear radius.

The current theoretical accuracy corresponds to an uncertainty of about $0.010$ fm.
Upon completion of the ongoing calculation of all one-electron two-loop QED effects \cite{volkov:25:tobe}, the theoretical accuracy will improve by roughly a factor of three; see Ref.~\cite{yerokhin:25:lilike} for a summary of present uncertainties.
In the more distant future, the theoretical limitations will likely be set by the nuclear polarization correction, whose uncertainty for $^{208}$Pb is on the order of 0.001 eV in modern calculations \cite{valuev:24}.

We therefore conclude that absolute determinations of nuclear radii from $2p$-$2s$ transitions of Li-like ions with an accuracy of about $0.005$\,fm appear feasible in the near future, limited by experimental transition energies rather than nuclear-structure effects, as is the case for muonic atoms.

Apart from determinations of nuclear radii, the present study will be useful for improved tests of bound-state QED in high-$Z$ ions.
The uncertainty due to the assumed nuclear-charge model contributes to the theoretical error of transition energies in these systems~\cite{yerokhin:15:Hlike,yerokhin:25:lilike}.
The developed approach allows us to eliminate the nuclear-model uncertainty.
This requires knowledge of the nuclear moment ratios $V_{24}$ and $V_{26}$.
For isotopes without available electron-scattering data, these ratios can nowadays be obtained by ab initio nuclear calculations~\cite{Beyer:25, miyagi2025ab}.
In the future, it would be desirable to produce a compilation of nuclear parameters that includes not only the rms charge radii, but also the ratios of the higher-order moments $V_{24}$ and $V_{26}$.

\appendix

\section{Nuclear models}
\label{sec:models}

Here we summarize the models of the nuclear charge distribution employed in this work.
Note that our choice of models is not primary guided by the goal of achieving a fully
realistic description of the nuclear charge distribution. Rather, we select
a set of models that (i) differ sufficiently from one another and (ii) are
flexible enough to allow fitting the lowest moments of a realistic charge distribution.
An additional criterion is that all models considered allow an analytical evaluation
of the charge distribution moments, which considerably simplifies our analysis.

The first class of models considered here are the Fermi models.
The standard two-parameter Fermi (2pF) distribution is
\begin{align}\label{eq:2pF}
\rho_{\rm 2pF}(r;c,a) = \frac{\rho_{0}}{1 + \exp[(r-c)/a]}\,,
\end{align}
where the normalization prefactor is given by
\begin{align}\label{eq:2pF:0}
\rho_{0}^{-1} = -8\pi a^3 {\rm Li}_3\big[-\exp(c/a)\big] \,,
\end{align}
with ${\rm Li}_n(x)$ being the polylogarithm function.
In practical calculations, the parameter $a$ is often fixed by $a_0 = 2.3/(4\ln3)$, which
leads to the one-parameter Fermi model,
\begin{align}\label{eq:1pF}
\rho_{\rm 1pF}(r;c) \equiv \rho_{\rm 2pF}(r;c,a_0)\,.
\end{align}
In the present work, we also use the three-parameter Fermi (3pF) model \cite{vries:87}
\begin{align}\label{eq:3pF}
\rho_{\rm 3pF}(r;c,a,w) = \rho_{0}\,
\frac{1+w r^2/c^2}{1 + \exp[(r-c)/a]}
\,.
\end{align}

The second class of models is based on the homogeneously charged sphere distribution. The corresponding
one-parameter (1pH) model is
\begin{align}\label{eq:1pH}
\rho_{\rm 1pH}(r;R) = \frac{3}{4\pi R^3}\,\theta(R-r)\,,
\end{align}
where $\theta(x)$ is the step function.
The parameter $R$ is connected to the rms radius by $r_C = \sqrt{3/5}R$.
Furthermore, we construct an $n$-parameter generalization of the 1pH distribution
as
\begin{align}\label{eq:npH}
\rho_{n{\rm pH}}(r;R_1\ldots R_n) = \sum_{i = 1}^n C_i\, \rho_{\rm 1pH}(r;R_i)\,,
\end{align}
where $C_i$'s are some fixed parameters with the condition $\sum_iC_i = 1$.
The choice of $C_i$ is, in principle, arbitrary, as long as it
allows for an adequate fitting of the nuclear-charge moments.

The third type of models is the Gaussian distribution. The one-parameter Gaussian (1pG) model is
\begin{align}\label{eq:1pG}
\rho_{\rm 1pG}(r;R) = \left(\frac{3}{2\pi R^2}\right)^{3/2}\,\exp\left(-\frac{3\,r^2}{2R^2}\right)\,,
\end{align}
with parameter $R$ connected with the rms radius by $r_C = R$.
The construction of an $n$-parameter generalization of the Gaussian model requires some
care.
A simple superposition of several 1pG distributions does not allow us to simultaneously
fit several realistic moments of the charge distribution while also ensuring that
the distribution remains positive for all $r > 0$.
We therefore choose to extend the 1pG distribution with one or several shifted Gaussian functions
introduced by Sick~\cite{sick:74}. This addition provided our anzatz enough flexibility
to fit the required moments of the charge distribution. The resulting
$n$-parameter Gaussian (npG) model is
\begin{align}\label{eq:npG}
\rho_{{\rm npG}}& \ (r;R_1\ldots R_n) = C_1\,\rho_{\rm 1pG}(r;R_1)
 \nonumber \\ &
+
\sum_{i = 2}^n C_i\, \frac1{2\pi^{3/2}\gamma(2R_i^2+\gamma^2)}
  \Big( e^{-\frac{(r-R_i)^2}{\gamma^2}} + e^{-\frac{(r+R_i)^2}{\gamma^2}}
 \Big)\,.
\end{align}
Here,
$C_i$ are some fixed parameters with $\sum_iC_i = 1$ and we fixed the parameter
$\gamma$ as $\gamma = \sqrt{2/3}\ 1.7$~fm~\cite{vries:87}.

The fourth type of models is the exponential distribution. The one-parameter exponential
model is given by
\begin{align}\label{eq:1pE}
\rho_{\rm 1pE}(r;R) = \frac{3\sqrt{3}}{\pi R^3}\,\exp\left(-\frac{2\sqrt{3}\,r}{R}\right)\,,
\end{align}
with the parameter $R$ connected to the rms radius by $r_C = R$.
Our generalization to the $n$-parameter model is constructed similarly to the
Gaussian case,
\begin{align}\label{eq:npE}
\rho_{{\rm npE}}& \ (r;R_1\ldots R_n) = C_1\,\rho_{\rm 1pE}(r;R_1)
 \nonumber \\ &
+
\sum_{i = 2}^n C_i\, \frac1{2\pi^{3/2}\gamma(2R_i^2+\gamma^2)}
  \Big( e^{-\frac{(r-R_i)^2}{\gamma^2}} + e^{-\frac{(r+R_i)^2}{\gamma^2}}
 \Big)\,.
\end{align}


\end{document}